\documentclass[12pt]{article}
\usepackage{epsfig}
\input epsf.sty
\newcommand{\la}[1]{\label{#1}}
 
\newcommand{\be}{\begin{equation}}
\newcommand{\ee}{\end{equation}}
\newcommand{\ba}{\begin{eqnarray}}
\newcommand{\ea}{\end{eqnarray}}
\newcommand{\bi}{\begin{itemize}}
\newcommand{\ei}{\end{itemize}}
\newcommand{\rmi}[1]{{\mbox{\scriptsize #1}}}
\newcommand{\nr}[1]{(\ref{#1})}
\newcommand{\tr}{{\rm Tr\,}}

\newcommand{\fr}[2]{{\frac{#1}{#2}}}
\newcommand{\msbar}{\overline{\mbox{\rm MS}}}

\renewcommand{\vec}[1]{{\bf #1}}

\newcommand{\RR}{{\rm I\kern -.2em  R}} 
\newcommand{\eq}{Eq.~}
\newcommand{\eqs}{Eqs.~}
\newcommand{\fig}{Fig.~}

\newcommand{\se}{Sec.~}

\def\lsi{\raise0.3ex\hbox{$<$\kern-0.75em\raise-1.1ex\hbox{$\sim$}}}
\def\gsi{\raise0.3ex\hbox{$>$\kern-0.75em\raise-1.1ex\hbox{$\sim$}}}
\newcommand{\lsim}{\mathop{\lsi}}
\newcommand{\gsim}{\mathop{\gsi}}

\makeatletter
\renewcommand\section{\@startsection {section}{1}{\z@}%
                                   {-5.5ex \@plus -1ex \@minus -.2ex}
                                   {2.3ex \@plus.2ex}%
                                   {\normalfont\large\bfseries}}
\renewcommand\subsection{\@startsection{subsection}{2}{\z@}%
                                     {-3.25ex\@plus -1ex \@minus -.2ex}%
                                     {1.5ex \@plus .2ex}%
                                     {\normalfont\normalsize\bfseries}}
\renewcommand\thesection {\@arabic\c@section}
\renewcommand\thesubsection   {\thesection.\@arabic\c@subsection}
\renewcommand{\@seccntformat}[1]{%
\csname the#1\endcsname.\hspace{1.0em}}
\makeatother

\begin{document}
 
\begin{titlepage}
\begin{flushright}
CERN-TH/2001-101\\
hep-ph/0104031\\
\end{flushright}
\begin{centering}
\vfill
 
{\bf A REMARK ON HIGHER DIMENSION INDUCED\\ DOMAIN WALL DEFECTS IN OUR WORLD}

\vspace{0.8cm}
 
C.P. Korthals Altes$^{\rm a,}$\footnote{chris.korthal-altes@cpt.univ-mrs.fr},
M. Laine$^{\rm b,}$\footnote{mikko.laine@cern.ch} 

\vspace{0.3cm}
{\em $^{\rm a}$%
Centre Physique Theorique, CNRS, 
Case 907, Luminy, F-13288 Marseille, France\\}
\vspace{0.3cm}
{\em $^{\rm b}$%
Theory Division, CERN, CH-1211 Geneva 23,
Switzerland\\}

\vspace*{0.8cm}
 
\end{centering}
 
\noindent
There has been recent interest in new types of topological 
defects arising in models with compact extra dimensions. We discuss 
in this context the old statement that if only SU($N$) gauge fields 
and adjoint matter live in the bulk, and the coupling is weak, then 
the theory possesses a spontaneously broken global Z($N$) symmetry, 
with associated domain wall defects in four dimensions. We discuss
the behaviour of this symmetry at high temperatures. We argue that 
the symmetry gets restored, so that cosmological domain wall 
production could be used to constrain such models. 
\vfill
\noindent
 

\vspace*{1cm}
 
\noindent
CERN-TH/2001-101\\
May 2001 

\vfill

\end{titlepage}

\section{Introduction}

There has been recent interest in models where our 
four-dimensional (4d) world 
is a thin brane embedded in a number of other, compact dimensions. 
Apparently
there are many alternatives as to which fields are confined to the brane 
and which are free to propagate in the bulk. We shall here consider 
the case that some non-Abelian gauge fields live in the 
bulk~(\cite{ia} and variants thereof;  
for a review, see~\cite{ddg}). For concreteness, 
we mostly concentrate on the case of a single extra dimension. 
We then wish to ask what kind of topological defects in our 4d world such 
a situation may lead to. 

Let us recall that 
the Standard Model of particle physics does not support 
any stable topological defects, but many extensions thereof do. 
For instance, Grand Unified Theories may predict the existence
of monopoles. It has recently been pointed out that the existence
of compact extra dimensions could also lead to various types of 
topological defects on our 4d brane~\cite{dks}. 

The existence of topological defects can clearly have important 
implications for cosmology. Conversely, the fact that 
none have been observed directly or indirectly, places a number of
strong constraints on theories beyond the Standard Model. 

In this note we discuss one special topological defect possibly 
arising in the aforementioned models with compact extra dimensions.
The same mechanism has previously been addressed in~\cite{bk},
and a somewhat analogous with two compact dimensions in~\cite{nhs}.   

\section{Z($N$) symmetry}

It is well known that when SU($N$) gauge fields are compactified
with periodic boundary conditions, the system develops a
global Z($N$) symmetry, which is spontaneously broken for a small 
coupling~\cite{gh,gpy,nw}. This leads to the existence of domain walls. 
Let us briefly recall the argument. 

We consider the simplest case of a flat periodic extra dimension, 
$y = 0...R$. As far as we can see, the statement holds also for instance
for an orbifold, $S_1/Z_2$, with the difference that the number of 
effective gauge degrees of freedom is halved, since the 
extra symmetry removes the imaginary modes from  
the Fourier decomposition of a real field. Let us denote $M = R^{-1}$.
SU($N$) gauge theory in dimensions higher than 3+1 is not renormalizable, 
but for concreteness we assume that there is another cutoff scale 
$\Lambda \gg M$, up to which the dominant effects come just 
from the Yang-Mills Lagrangian. The observables we 
compute with this theory are in any case finite. 

The partition function of the system is now  
\be
Z = \int_{\rmi{b.c.}} {\cal D} A_\mu 
\exp\Bigl(-\int_0^R dy \int d^{d-1}x 
\,\fr12 \tr F_{\mu\nu}^2 + ...\Bigr),
\la{Z0}
\ee
where the boundary conditions are periodic,
$A_\mu(x,0) = A_\mu(x,R)$, and   
higher dimensional operators are ignored. 
Our convention in the following is that
$D_\mu = \partial_\mu + i g A_\mu$.

Consider now field transformations $U(x,y)$, $x=(x_0,...,x_{d-2})$, 
which look locally just 
like gauge transformations, but have the property that 
\be
U(x,R) = z U(x,0), \la{trans}
\ee
where $z = \exp(i 2\pi k/N)\in$ Z($N$), with $k=0,...,N-1$.
The Lagrangian is clearly invariant under these transformation.
In addition, considering transformations where $U$ factorizes into 
the form on the RHS of \eq\nr{trans} also for $y\in(R-\epsilon,R)$,
the fields, 
\be
A_\mu \to 
A_\mu' = U A_\mu U^\dagger + \frac{1}{ig} U \partial_\mu U^\dagger, 
\ee
remain periodic.
Thus, $z$ in \eq\nr{trans} represents a global symmetry of the theory. 

Fundamentally charged fields in the bulk, 
on the other hand, spoil the symmetry: 
after the transformation $\Phi \to \Phi' = U \Phi$,
$\Phi'$ would not respect the original boundary conditions.
Which of the Z($N$) vacua becomes the global one, 
depends on the original boundary conditions~\cite{nw,hosotani}.
There might be ways of avoiding this explicit symmetry
breaking, however. We might for instance imagine that most fundamentally 
charged fields are strictly confined to a brane as in some
orbifold compactifications~(see, e.g., \cite{ddg}), so that 
they are insensitive to the extra dimension (i.e., do not 
couple to the corresponding covariant derivative), and the 
symmetry breaking effects are suppressed. Or we could imagine
that the gauge group in question is not one of the Standard Model 
ones, and only feels adjoint matter. Later on, we shall also 
return briefly to the case where fundamental charges 
do cause an explicit symmetry breaking. 

Now, all local gauge invariant operators are invariant 
under the transformation in \eq\nr{trans}. 
There is a non-local gauge invariant operator which is 
not invariant, however: 
\be
P(x) = \frac{1}{N} \tr {\cal P} \exp( ig\int_0^R dy A_y(x,y)), \quad
P(x) \to P'(x) = z^* P(x). \la{Px}
\ee 
Because $P(x)$ is a non-local operator, 
its absolute value does not have a meaningful
continuum limit. The phase factor of $\langle P(x)\rangle$, however,  
denoted by $z$ in the following, is assumed to be physical, 
and acts as an order parameter for the symmetry.
 
This symmetry takes a more familiar form for instance 
in the set of gauges $\partial_y A_y = 0$. 
Then it just corresponds to a shift in $A_y$. 
As we see from \eq\nr{Px}, the vacuum
with $z=1$ is obtained for $A_y = 0$, while vacua with
$z = \exp(i 2\pi k/N)$ for instance with 
$g A_y = (2 \pi k/(R N)) \times \mathop{\mbox{diag}}(1,1,...,1-N)$. 

Despite the somewhat abstract nature of this symmetry, 
it does lead to domain walls carrying a finite energy density in 4d, 
in case the symmetry is broken\footnote{For completeness 
let us note that these Z($N$) domain walls are not 
related to those found in 4d supersymmetric Yang-Mills theory
(for recent reviews, see~\cite{chmm}).}. 
The reason is that once quantum corrections are taken into account, 
the effective potential for $A_y$, or more precisely the
constrained effective action for $P(x)$, develops 
barriers between the minima corresponding to 
$\arg \langle P \rangle = 2\pi k/N, k = 0,...,N-1$. 
The computation of such quantum corrections,
and thus this statement, is reliable for a weak coupling. 

To proceed, we may parameterise $gA_y = C = \mathop{\mbox{diag}}
(C_1,C_2, ..., C_N)$, $\sum_i C_i = 0$, 
so that $P \sim (1/N) \tr \exp(i R C)$.  
In principle we want to compute the constrained
effective action
\be
e^{-S_\rmi{eff}[C(x)]} \sim 
\biggl\langle
\Pi_x \delta\Bigl(\frac{1}{N} \tr e^{i R C(x)} -  P(x) \Bigr)
\biggr\rangle.   
\ee
At 1-loop level, and for 
$x$-independent configurations of $C$, 
this however reduces simply to the standard 
effective potential $V_\rmi{eff}(C)$: 
$S_\rmi{eff}\to R L^{d-1} V_\rmi{eff}(C)$, 
where $L^{d-1}$ is the volume in $d-1$ dimensions of extent $L$.
By construction, $V_\rmi{eff}$ is gauge independent. 

In order to show the result for $V_\rmi{eff}(C)$, 
we denote $C_{ij} = (C_i - C_j)/(2 \pi M)$.
Adding together the contributions 
from the gauge bosons and the ghosts
in the background of $C$, the result
can be written in the form~\cite{gpy,nw} 
\ba
V_\rmi{eff}(C) - V_\rmi{eff}(0) & = & \sum_{i\neq j} 
\Bigl[ U(C_{ij}) - U(0) \Bigr], \la{veff} \\
U(C_{ij}) - U(0) & = & 
(d/2-1) M \sum_{l = -\infty}^{\infty} 
\int \frac{d^{d-1} p}{(2\pi)^{d-1}} \ln
\frac{p^2 + (2\pi M)^2 (l + C_{ij})^2}{p^2 + (2\pi M)^2 l^2}. \la{U}
\ea
We write the logarithm in a heat kernel form 
$\ln(a/a_0) = \int_0^\infty (ds/s) (e^{-a_0 s} - e^{-a s})$ and 
use the Poisson summation formula
\be
\sum_{l=-\infty}^{\infty} e^{-s (2\pi M)^2 (l + C_{ij})^2} = 
\frac{1}{2\pi M}\Bigl(\frac{\pi}{s}\Bigr)^{1/2}
\sum_{l=-\infty}^{\infty} e^{-l^2/(4 s M^2) + i 2\pi l C_{ij}} , 
\ee
obtained after the $x$-integration from 
$\sum_l f(l) = 
\int dx 
\sum_l 
e^{i 2\pi l x} f(x).$
The $s$-integration can also be carried out, to arrive at 
\be
U(C_{ij})-U(0)=2(d-2)M^d \frac{\Gamma(\frac{d}{2})}{\pi^{\frac{d}{2}}}
\sum_{l=1}^{\infty}{1\over{l^d}}\sin^2 (\pi l C_{ij}).
\label{eq:zerotemp}
\ee
This contribution vanishes at $C_{ij} = 0 \mathop{\mbox{mod}} 1$, 
otherwise it is positive, representing thus a barrier between the 
different degenerate minima.

To compute the energy density of the domain wall
interpolating between two different degenerate minima, 
derivative terms are needed too in $S_\rmi{eff}[C]$. 
At leading order, it is enough to keep the tree-level 
kinetic terms~\cite{bgap}. This leads to the action
\be
S_\rmi{eff}[C] = R \int d^{d-1} x \,
\biggl[
\frac{1}{g^2} \sum_{\mu = 0}^{d-2} \tr (\partial_\mu C)^2  + 
V_\rmi{eff}(C) 
\biggr].  \la{action}
\ee

Let us now consider two adjacent minima, 
$\arg \langle P\rangle = 2\pi k/N, k=0,1$.
The extremal path between $k=0,1$ is given by 
$C = (2\pi M q/N)\times \mathop{\mbox{diag}}(1,1,...,1-N)$, 
where $0\le q \le 1$~\cite{bgap,ga}. 
We minimise $S_\rmi{eff}[C]$ from 
\eq\nr{action} with these boundary conditions, 
assuming a planar symmetry.  
Converting the integral over the coordinate across the domain wall into 
one over $q$ with the standard procedure,   
we get for the energy per unit area 
\be
\sigma_1 = \frac{4\pi (N-1)}{g\sqrt{N} } 
\int_0^1 dq \sqrt{2\Bigl[U(q) - U(0)\Bigr]}. \la{sigma}
\ee
We are not aware of a general analytic answer for this integral,
given $U$ in \eq\nr{eq:zerotemp}.

We close this section with a remark. In asymptotically
free theories, the effective coupling $g$ is guaranteed 
to be small inside the wall (see below) but starts to 
grow in the wings and will be large outside. This makes 
for instance $U(0)$ uncomputable in confining theories. 
Nevertheless, the energy per unit area turns out to admit 
a perturbative expansion, since it is an integral over 
the profile and is dominated by the inside region. This 
infrared insensitivity has been demonstrated explicitly 
at next-to-leading order for $d=4$~\cite{bgap,ga}, 
and we shall return to the case $d=5$ below.

\section{4d at zero temperature}
\la{4d}

Because many explicit results are available, 
we next specialise briefly in a (2+1)d toy world, 
with a compact 4th dimension. In Euclidian spacetime, this situation
corresponds formally to a 4d field theory at a finite temperature, 
and has thus been thoroughly studied, although with a different
physical interpretation. 

In fact, the first study was by 't Hooft~\cite{gh}, with 
yet another language. He introduced gauge invariant electric
and magnetic fluxes. Then the statement is that if we 
consider the magnetic flux in a spatial direction (with extent $L_2$) 
orthogonal to the compact one, then for large $R$  
the extra energy related to the
flux is $E_m \sim \rho L_2 \exp(-\rho L_1 R)$, 
where $L_1$ is the other of the spatial dimensions, 
and $\rho$ is the electric string tension of the 
confining phase of pure 4d Yang-Mills. For small $R$, 
on the other hand, the effective coupling $g(2\pi M)$ 
is weak and we find $E_m \sim \sigma_1 L_2$, where 
$\sigma_1\sim 1/R^2$ is given in \eq\nr{sigma}.

Let us now recall more precisely
when the perturbative computation is reliable.  
In the 4d case, the original 
gauge coupling $g_4^2$ is dimensionless, while 
that of the 3d world is to 
leading order $g_3^2 = g_4^2 M$. If we compute
the effective potential, the only dimensionful 
parameters entering at 1-loop level are multiples
of $2\pi M$. Thus we assume that the effective
dimensionless expansion parameter related to the
computation is 
\be
\epsilon \sim \frac{\alpha}{\pi} \sim \frac{g_4^2(2\pi M)}{4\pi^2}. \la{eps}
\ee 
For $M$ larger than the scale parameter in $g_4^2$, 
the expansion parameter is thus small. In the 
finite temperature case, this corresponds to the 
deconfined phase of QCD. 

Then, the sum in \eq\nr{eq:zerotemp} can be explicitly
carried out, leading to a Bernoulli polynomial $B_4$~\cite{gpy,nw}:
\be
U(C_{ij})- U(0) = 
\fr23 \pi^2 M^4 
[C_{ij} \mathop{\mbox{mod}} 1]^2(1 - [C_{ij} \mathop{\mbox{mod}} 1])^2.
\ee
Consequently~\cite{bgap,ga}, 
\be
\sigma_1 = \frac{4\pi^2(N-1)}{3\sqrt{3N}}\frac{M^2}{g_4}, \quad
\sigma_k =  \sigma_1 \frac{k(N-k)}{N-1}.
\ee
Here $k > 1$, relevant for $N > 3$, corresponds to 
a profile between $z=1$ and $z=\exp(i2\pi k/N)$.
Next-to-leading order (2-loop) corrections are
also known~\cite{bgap,ga,ka1}. This domain wall energy density
has effectively also been measured with lattice simulations
(\cite{lattice} and references therein), although again with 
a different interpretation. 

\section{5d at zero temperature}

In the physically more interesting
case of a 5d theory, the original 
gauge coupling $g_5^2$ has the dimension GeV$^{-1}$.
Let us define $g_4^2 = g_5^2 M$, and assume that the 
theory can be regularised such that a perturbative 
expansion in $g_4^2$ makes sense.  
Then, the only dimensionful mass
parameters entering at 1-loop level are multiples
of $2\pi M$. Thus the effective
expansion parameter related to the
computation is essentially the same as in \eq\nr{eps}, 
as long as we are inside the wall.
For $M$ in the TeV range or so, but smaller 
than the cutoff of the 5d theory, a weak coupling 
computation should thus be reliable. 

Now we are not able to evaluate \eq\nr{sigma} analytically, however. 
We find 
\be
\sigma_1 = 0.622988 \times 4\pi \frac{N-1}{\sqrt{N}} \frac{M^{\fr52}}{g_5},
\quad
\sigma_k =  \sigma_1 \frac{k(N-k)}{N-1},
\la{s1}
\ee
where the first numerical equality is 
in accordance with~\cite{bk}, and the latter part, 
indicating that two different 
domain walls tend to attract each other, follows from~\cite{ga}. 
Expressed in terms of $g_4$, 
the dimensionful combination here is $M^{\fr52}/g_5 = M^3/g_4$.

\section{5d at finite temperature}
\la{finiteT}

\begin{figure}[t]

\centerline{\epsfxsize=6cm\hspace*{0cm}\epsfbox{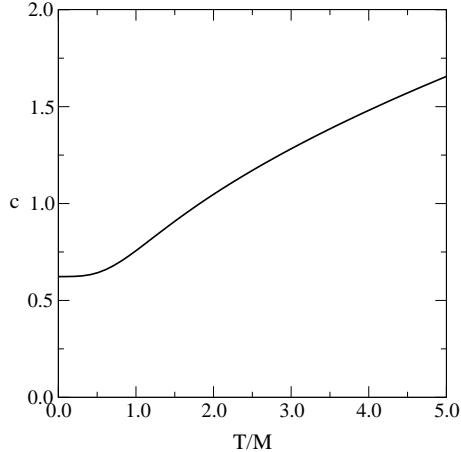}}

\caption[a]{The coefficient $c$ 
in the domain wall energy density (cf.\ \eq\nr{sigma}),
$\sigma_1 \equiv c \times 4\pi(N-1)M^{\fr52}/(g_5 \sqrt{N})$,
as a function of $T/M$. We have used \eqs\nr{sigma}, 
\nr{veffT}. At large $T/M$ the behaviour is $\sim (T/M)^{\fr12}$.}

\la{fig:sigma}
\end{figure}

Let us finally consider the 5d system at finite physical temperature $T$. 
Some finite $T$ effects have previously been discussed for an 
analogous Abelian 3d system in~\cite{ms}. 

For reference, let us recall what happens at finite temperatures
for a standard Z(2) symmetry breaking model such as a 4d real scalar
field with the potential $V(\phi)\sim \lambda (\phi^2-v^2)^2$. The
dominant temperature correction is now
$\delta V(\phi) \sim T^2\phi^2$. Thus, 
if we just plot the effective potential as a function of $\phi$, 
the barrier between the two minima gets smaller
as $T$ goes up, and vanishes at $T = T_c \sim v$. The surface 
energy density, obtained by 
the analogue of \eq\nr{sigma}, decreases with
increasing $T$. Finally, the physical correlation length, 
$\sim 1/\sqrt{V''(\phi)}$, increases,
and diverges at $T=T_c$.
 
To see whether the same happens here, we compute $V_\rmi{eff}(C)$
to 1-loop order at a finite temperature $T$.
According to the standard procedure, 
a path integral formulation can be obtained simply by 
making the Euclidian time direction $\tau$ periodic, 
with the period $1/T$. Then the integration measure 
changes as
\be
\int\frac{d^{d-1}p}{(2\pi)^{d-1}} f(p_\tau,\vec{p}) \to 
T \sum_m \int\frac{d^{d-2}p}{(2\pi)^{d-2}} f(2\pi T m,\vec{p}).
\ee
Correspondingly, the effective potential for $C$ is modified. 
Using again the Poisson transformed heat kernel expression
for \eq\nr{U}, we obtain 
\ba
U(C_{ij})-U(0) = 
2(d-2)M^d \frac{\Gamma(\frac{d}{2})}{\pi^{\frac{d}{2}}}
\sum_{l=1}^\infty \sin^2(\pi l C_{ij})
\biggl[\frac{1}{l^d}+ \sum_{m=1}^\infty 
\frac{2}{(l^2+m^2/\rho^{2})^{\frac{d}{2}}}
\biggr],  \la{veffT}
\ea
where $\rho = T/M$. The $T=0$ contribution 
is the first term inside the square brackets. 

Curiously enough, we now note that each term in 
the series representing the change with a finite $T$ 
is {\em positive}. The value of the sum inside the 
square brackets in \eq\nr{veffT} grows like $\sim\rho$
at large $\rho$. Thus, introducing a temperature
{\em increases the barrier between the minima}. Correspondingly, 
the domain wall energy density increases as $\sim \rho^{\fr12}$ , 
as shown in \fig\ref{fig:sigma}. The second derivative 
at the minimum also goes as  $\sim\rho$, 
meaning that the correlation length
(and the width of the wall)
decrease as $\sim\rho^{-\fr12}$. 

Could this mean that the symmetry is not restored at high 
temperatures? In principle this is not excluded: there are well studied 
examples even of systems with inverse symmetry breaking~\cite{inv}. 
It seems to us, however, that at very high temperatures $T\gg M$,
the Z($N$) symmetry discussed here does get restored,
despite the behaviour found in the previous paragraph.  
This also means the perturbation theory breaks down in such a case, 
as we shall indicate presently. 

The argument is the old one~\cite{rp}. 
Indeed, our order parameter $\arg \langle P(x) \rangle$
lives in 3+1 dimensions, so we may think in terms of usual 
statistical mechanics. Then, creating a bubble of some Z($N$) phase 
with a surface area $A$ is Boltzmann 
suppressed by $\exp(-\sigma_1 A/T)$. But there are a lot of 
such configurations, and the corresponding entropy factor 
goes as $\exp(\Lambda^2 A)$, where $\Lambda$ is 
some ultraviolet cutoff~\cite{ambjorn}. We may expect $\Lambda^{-1}$
to be given by the width of the wall. According to the 1-loop
discussion above, $\sigma_1 \sim (M^3/g_4)\rho^{\fr12}$ 
and $\Lambda^2 \sim (g_4 M)^2 \rho$, where $\rho = T/M$. 
The symmetry should get restored when these 
two opposing factors compensate for each other
such that the exponent is of order unity~\cite{rp}. 
This\footnote{We do not consider here the running of the 
effective coupling with $T$, since we shall argue presently
that radiative corrections are parametrically subdominant 
for $T < T_c$.} leads to $T_c \sim M/g_4^2$.

There is another way of obtaining the same result. 
Imagine that we construct a dimensionally reduced effective theory 
by integrating out the non-zero Kaluza-Klein modes in the $y$-direction. 
The effective theory is 4d SU($N$) $+$ adjoint scalar matter, and its 
physics represents that of a broken Z($N$) phase of the 5d theory
(for a discussion in the 4d context, see~\cite{adjoint}).
Thus the symmetry is broken, as long as dimensional reduction 
is accurate. It can lose its accuracy only if the power
suppressed higher order operators truncated from the 4d 
effective theory become as important as the 
renormalizable ones, i.e., if the lightest dynamical mass
scales $m$ within the effective theory satisfy $m\gg 2\pi M$.
At finite temperatures, the confinement scale of 4d non-Abelian
gauge theory is $g_4^2(2\pi T) T$~\cite{linde,gpy}, 
where the approximate (say, $\msbar$) scale of $g_4^2$
has also been shown; the precise value has no significance here, 
even if it were $\propto 2\pi M$. Thus, we may 
expect a breakdown and a transition at $g_4^2(2\pi T_c) T_c \sim 2\pi M$, 
parametrically just as above. The phase diagram of the system 
in this language is illustrated in \fig\ref{fig:phasediag}.

Finally, let us note that at the transition point, perturbation
theory breaks down. Indeed, 
loops involving the temporal Matsubara zero modes 
obtain a dimensionless expansion parameter  
$\gsim g_5^2 T/(2\pi) \sim g_4^2 T/(2\pi M)$.
This is parametrically small at $T < T_c$, but becomes
of order unity at $T \sim T_c$. 
Thus the effective potential as well as $\sigma_1$ and the width of the wall 
computed above, need no longer be reliable, 
and $\sigma_1$ could in principle even vanish at $T=T_c$. 

\begin{figure}[t]

\centerline{\epsfxsize=7cm\hspace*{0cm}\epsfbox{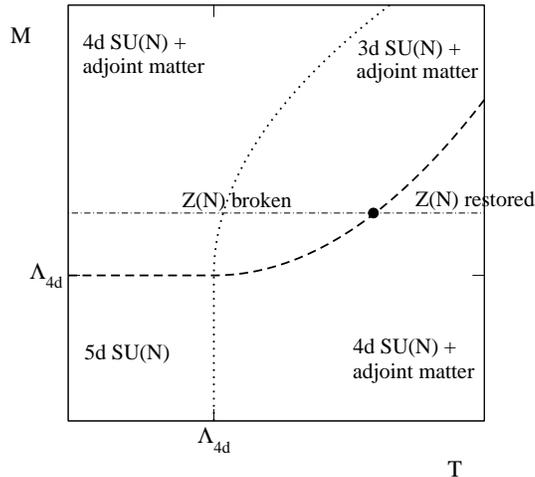}}

\caption[a]{A schematic phase diagram, together with the  
low-energy effective theories in different regions. The Z($N$) symmetry
we have discussed is broken above the dashed line. To the 
right of the dotted line, another Z($N$) symmetry, related to a Polyakov
loop in the Euclidean (finite-$T$) time direction is broken;  
its physical interpretation is however somewhat ambiguous, 
so we do not dwell on it here. Our argument follows the horizontal 
dash-dotted line, with the blob indicating the phase transition.}

\la{fig:phasediag}
\end{figure}

While we consider the arguments presented credible, they of course 
do not constitute a proof for the existence of a phase transition, 
because of the breakdown of perturbation theory. 
It would be interesting to study the issue non-perturbatively, 
at least for the 4d model discussed in \se\ref{4d}, where 
a similar reasoning can be carried out, leading to 
the same parametric estimate for $T_c$.  

\section{Cosmology}

We end with a brief cosmological consideration. 
It is well known that if there is a phase transition
after inflation where a discrete symmetry gets broken, domain
walls are generically produced~\cite{z}.
They would disappear slowly because the finite speed
of light imposes causality, and carry a lot of energy. In fact, they would 
soon be the dominant energy component, and
conflict with the observed energy density fluctuations in 
the cosmic microwave background, unless 
$\sigma \lsim (1 \mbox{ MeV})^3$~\cite{z,vs}.

The constraint is weaker if the walls are not absolutely 
stable. This can happen for instance if the Z($N$) symmetry is not 
exact. Such is the case with some fundamental
matter in the bulk; the minima corresponding to $z\neq 1$
are then lifted or lowered, depending on the boundary
conditions, but they may remain metastable~\cite{nw,hosotani}. 
In conventional cosmology wall domination is avoided 
if $\Delta \epsilon \gg \sigma^3/m_\rmi{Pl}^2$, 
where $m_\rmi{Pl} = 1.2\times 10^{19}$ GeV, 
and $\Delta \epsilon$ is the energy density excess 
in the metastable state~\cite{vs}. In the present 
case $\Delta\epsilon \sim M^4, \sigma \sim M^3/g_4$, 
so that we arrive at a rather weak constraint $M \ll g_4 m_\rmi{Pl}$.
The situation could change in alternative
cosmologies probably more relevant in the presence
of large extra dimensions. 

Looking back at the first paragraph of this Section, we thus see
that if the Z($N$) symmetry gets restored at 
high $T$, as we have suggested in \se\ref{finiteT}, 
then either particle physics models with 
only adjoint matter in the bulk, 
or cosmological models with temperatures above
$\sim M/g_4^2$, can be excluded. 
Some fundamental matter in the bulk 
relaxes these constraints. There are many other 
constraints, of course, which need to be met as well; 
see, e.g.,~\cite{ddg} and references therein. 

\section{Conclusions}

It is well known that if SU($N$) gauge fields and adjoint matter
are compactified in an extra spacelike direction, then the theory
develops a global Z($N$) symmetry. If the effective coupling is weak, 
the symmetry is broken, resulting in domain wall configurations.
These are in principle visible in the remaining flat spacetime. 

In order for this observation 
to be physically relevant, at least two questions have
to be answered. The first is, what is the effect of fundamentally
charged matter? It is known that if it propagates in the extra 
dimension, it breaks the Z($N$) symmetry explicitly. One possible
way to avoid this could be to confine fundamentally
charged matter strictly to a set of branes, another to 
consider gauge fields not belonging to the Standard Model
and only interacting with adjoint matter.

The second question is, would such domain walls have any 
practical significance? They have a huge energy density, and 
thus are not produced under any normal circumstances.  But if the
Z($N$) symmetry gets restored at high temperatures, they could
be produced in the Early Universe, which would give strong constraints 
on particle physics models of this type, or on cosmology. We have discussed 
the behaviour of  the Z($N$) symmetry at high temperatures, and shown
that at strict 1-loop level the symmetry does {\em not} get restored. 
We have argued however that non-perturbatively it does get restored at 
temperatures of order $M/g_4^2$, where $M$ is the mass scale
related to the compact direction. Under these circumstances, 
either particle physics models with 
only adjoint matter in the bulk, or 
cosmologies with temperatures above $\sim M/g_4^2$, 
can be excluded, because cosmologically produced domain 
walls would induce energy density 
fluctuations much larger than those observed
in the cosmic microwave background. 


\section*{Acknowledgements}

We thank A.~Kovner and M.~Shaposhnikov for valuable discussions. 
This work was partly supported by the TMR network {\em Finite
Temperature Phase Transitions in Particle Physics}, EU contract no.\ 
FMRX-CT97-0122, and by the RTN network {\em Supersymmetry and 
the Early Universe}, EU contract no.\ HPRN-CT-2000-00152.

\end{document}